\documentclass[conference]{IEEEtran}
\IEEEoverridecommandlockouts
\usepackage{cite}
\usepackage{amsmath,amssymb,amsfonts}
\usepackage{algorithmic}
\usepackage{graphicx}
\usepackage{textcomp}
\usepackage{xcolor}
\usepackage{url}
\usepackage{float}
\usepackage{subfig}
\usepackage{caption}
\usepackage{diagbox}
\usepackage{array}
\usepackage{booktabs}
\usepackage{hyperref}
\usepackage{multirow}
\usepackage{float}
\usepackage[misc]{ifsym} 
\def\tablename{Table}
\def\BibTeX{{\rm B\kern-.05em{\sc i\kern-.025em b}\kern-.08em
    T\kern-.1667em\lower.7ex\hbox{E}\kern-.125emX}}
\begin{document}

\title{Towards Strengthening Deep Learning-based Side Channel Attacks with \textit{Mixup}}

\author{\IEEEauthorblockN{Zhimin Luo$^{\textrm{\Letter}}$, Mengce Zheng, Ping Wang, Minhui Jin, Jiajia Zhang and Honggang Hu$^{\textrm{\Letter}}$}
\IEEEauthorblockA{\textit{Key Laboratory of Electromagnetic Space Information, CAS} \\
\textit{University of Science and Technology of China}, Hefei, China \\
Email: zmluo@mail.ustc.edu.cn, hghu2005@ustc.edu.cn}
}

\maketitle

\begin{abstract}
In recent years, various deep learning techniques have been exploited in side channel attacks, with the anticipation of obtaining more appreciable attack results. Most of them concentrate on improving network architectures or putting forward novel algorithms, assuming that there are adequate profiling traces available to train an appropriate neural network. However, in practical scenarios, profiling traces are probably insufficient, which makes the network learn deficiently and compromises attack performance. 

In this paper, we investigate a kind of data augmentation technique, called \textit{mixup}, and first propose to exploit it in deep-learning based side channel attacks, for the purpose of expanding the profiling set and facilitating the chances of mounting a successful attack. We perform Correlation Power Analysis for generated traces and original traces, and discover that there exists consistency between them regarding leakage information. Our experiments show that \textit{mixup} is truly capable of enhancing attack performance especially for insufficient profiling traces. Specifically, when the size of the training set is decreased to 30\% of the original set, \textit{mixup} can significantly reduce acquired attacking traces. We test three \textit{mixup} parameter values and conclude that generally all of them can bring about improvements. Besides, we compare three leakage models and unexpectedly find that least significant bit model, which is less frequently used in previous works, actually surpasses prevalent identity model and hamming weight model in terms of attack results.
\end{abstract}


\section{Introduction}
\label{sec1}

Since Paul Kocher proposed timing attacks to recover secret information in 1996 \cite{10.1007/3-540-68697-5_9}, side channel attacks (SCA) have been an immense threat to the security of cryptosystems. They exploit physical leakages of embedded devices, where a cryptographic algorithm is implemented. During every execution, the implementation manipulates sensitive variables which depend on a portion of public knowledge (e.g. plaintext) and some chunk of secret data (e.g. key). In SCA, several prevalently utilized physical leakages incorporate timing \cite{10.1007/3-540-68697-5_9}, power consumption \cite{kocher1999differential} and electromagnetic emanation \cite{10.1007/3-540-36400-5_4}. 

Ordinarily, SCA can be classified into two categories: non-profiled attacks and profiled attacks. The former directly  executes statistical calculations upon measurements of the target device with respect to a hypothetical leakage model. Commonplace instances comprise Differential Power Analysis (DPA) \cite{kocher1999differential}, Correlation Power Analysis (CPA) \cite{2004Correlation} and Mutual Information Analysis (MIA) \cite{2008Mutual}. Nevertheless, profiled attacks exhibit more considerable potential to break the target implementation. In profiled attacks, the attacker is presumed to be capable of thoroughly dominating a profiling device identical to the target device. The attacker can thereby construct a model to characterize the physical leakage and subsequently carry out the key retrieval process. Template Attacks (TA) \cite{chari2002template} and stochastic attacks \cite{Schindler2005A} are customary fashions in this domain.

Realistically, attackers regularly perform the preprocessing procedure to attain approving experimental results, in which case dimensionality reduction like Principal Component Analysis (PCA) \cite{Archambeau2006Template}, feature selection \cite{2019A} and noise reduction \cite{Wu_Picek_2020} beforehand are typical approaches. Recently, deep learning (DL)-based SCA has been advocated as an alternative to conventional SCA, among which multilayer perceptron (MLP) and convolutional neural networks (CNN) are most widely adopted. MLPs joint the features in the dense layers and accordingly imitate the effect of higher-order attacks, hence standing as a natural choice when dealing with masking countermeasures\cite{2016Breaking,Prouff2018StudyOD}. On the other hand, CNNs have the property of spatial invariance, bringing about intrinsic resilience against trace misalignment \cite{10.1007/978-3-319-66787-4_3,Picek_Heuser_Jovic_Bhasin_Regazzoni_2018,Timon_2019,2019AB,Zaid_Bossuet_Dassance_Habrard_Venelli_2020}. Furthermore, the internal architecture of CNNs make them adequate for automatical feature extraction from high-dimensional data, which eliminates the demand for preprocessing. Recent studies manifest that CNNs can yield at least equivalent performance compared to TA or even transcend TA \cite{Hettwer2019Profiled,Kim_Picek_Heuser_Bhasin_Hanjalic_2019,jin2020enhanced}.

Nonetheless, we notice that the aforementioned researches are based on the premise that profiling traces are ample to establish an appropriate model. While the down-to-earth plight is that we may encounter frustrations to be provided with inadequate measurements given time and resource constraints. This elicits an issue that how to boost the opportunities of launching a successful attack when confronted with deficient traces. Therefore, in this paper we propose to employ a data-augmentation technique, \textit{mixup}, in DL-based SCA to reinforce the power of existing insufficient traces and consequently enhance attack performance.

\subsection{Related Work}
\label{subsec1.1}

The great majority of work in the DL-based SCA community has sought to promote neural network architectures, preprocess traces or put forward new metrics, and then mounts a profiled SCA with an entire training set of some public datasets, anticipating the emergence of more efficient attacks. Reference \cite{Zaid_Bossuet_Habrard_Venelli_2019} interpreted the role of hyperparameters through some specific visualization techniques involving Weight Visualization, Gradient Visualization and Heatmap, and subsequently developed a methodology to systematically build well-performing neural networks. Reference \cite{Wouters_Arribas_Gierlichs_Preneel_2020} improved upon the
work from \cite{Zaid_Bossuet_Habrard_Venelli_2019}, where the authors indicated several misconceptions in the preceding paper and displayed how to obtain comparable attack performance with remarkably smaller network architectures. As illustrated in \cite{Wu_Picek_2020}, some types of hiding countermeasures were considered as noise and then the denoising autoencoder technique was used to remove that noise in traces. Whereas \cite{hettwer2020encoding} demonstrated and compared several methodologies to transform 1D power traces into 2D images. Reference \cite{Zhang_Zheng_Nan_Hu_Yu_2020} proposed a novel evaluation metric, called Cross Entropy Ratio (CER), that is closely connected to commonly employed side channel metrics and can be applied to evaluate the performance of DL models for SCA. They further adapted the CER metric to a new loss function to adopt in the training process, which can yield enhanced attack performance when dealing with imbalanced data. Another innovative loss function, Ranking Loss (RkL), was proposed by adapting the "Learning to Rank" approach in the Information Retrieval field to the side-channel context \cite{Zaid_Bossuet_Dassance_Habrard_Venelli_2020}, which helps prevent approximation and estimation errors, induced by the typical cross-entropy loss.

In \cite{Picek2019ProfilingSA}, Picek et al. emphasized the noticeable gap between hypothetical unbounded power and real finite power of an attacker. They considered a restricted setting where only limited measurements are available to the attacker for profiling and proposed a new framework, called the Restricted Attacker framework, to enforce attackers truly conduct the most powerful attack possible. Data augmentation (DA), which consists of a set of techniques that increase the size and quality of profiling sets, arises initially for computer vision tasks \cite{2019ABC} and has been applied to SCA recently. In \cite{10.1007/978-3-319-66787-4_3}, Cagli et al. augmented the training set by a simulated clock jitter effect. For the sake of eliminating the curse of data imbalance, \cite{Picek_Heuser_Jovic_Bhasin_Regazzoni_2018} employed The Synthetic Minority Oversampling Technique (SMOTE), which is a category of DA technique, to balance the class distribution of the training set. In this paper, we embrace another kind of DA technique---\textit{mixup}, an uncomplicated learning principle proposed to mitigate unacceptable behaviors such as memorization to adversarial examples in large deep neural networks \cite{DBLP:journals/corr/abs-1710-09412}, aiming at alleviating the trouble in a restricted position. 

\subsection{Contributions}
\label{subsec1.2}

Our contributions are chiefly summarized as follows.
\begin{enumerate}
	\item We investigate a DA technique, called \textit{mixup}, and first propose to exploit \textit{mixup} in practical SCA context to expand insufficient profiling set and enhance attack performance;
	\item We compare three leakage models and surprisingly find that least significant bit (LSB) model performs much better than another two models, which is unexpected since this model is almost ignored in related works;
	\item We apply \textit{mixup} to some publicly available datasets and test several \textit{mixup} parameter values and confirm that this technique can truly facilitate launching a successful DL-based SCA especially in restricted situations where profiling traces are inadequate.  
\end{enumerate}

\subsection{Organizations}
\label{subsec1.3}

The paper is organized as follows: Section \ref{sec2} provides the notations and preliminaries. Section  \ref{sec3} introduces the application of \textit{mixup} in DL-based profiled attacks and discusses the effect of chosen leakage models. In section \ref{sec4}, we display and analyse the experimental results on some datasets to illustrate this methodology's performance. Section \ref{sec5} summarizes the paper and gives insights on possible future work.

\section{Preliminaries}
\label{sec2}

\subsection{Notations}
\label{subsec2.1}

Throughout this paper, we use calligraphic letter $\mathcal{X}$ to denote sets, the corresponding upper-case letter $X$ to denote random variables (resp. random vectors $\mathbf{X}$) over $\mathcal{X}$, and the corresponding lower-case letter $x$ (resp. $\mathbf{x}$) to denote realizations of $X$ (resp. $\mathbf{X}$).  In SCA, the random variable is constructed as $X \in \mathbb{R}^{1 \times D}$, where $D$ denotes the dimension of the traces. The $i$-th entry of a vector $\mathbf{x}$ is denoted by $\mathbf{x}[i]$. We use $f$  to denote a cryptographic primitive which computes a sensitive intermediate $Z = f(P, K)$, where $P$ denotes a part of the public variable (e.g. plaintext or ciphertext) and $K$ denotes some chunk of the secret key that the attacker aims to retrieve. We denote $k^*$ as the true key of the cryptographic algorithm and $k$ as a key candidate that takes value from the keyspace $\mathcal{K}$.

\subsection{Deep Learning-Based Profiled Attacks}
\label{subsec2.2}

There exist two phases in a profiled attack: profiling phase and attacking phase, which correspond to the training stage and testing stage in a supervised learning task. In the profiling phase, the attacker generates a model $F:\mathbb{R}^{D} \rightarrow \mathbb{R}^{|\mathcal{Z}|}$ from a profiled set $\mathcal{T}=\left\{\left(\mathbf{t}_{0}, z_{0}\right), \ldots,\left(\mathbf{t}_{N_{p}-1}, z_{N_{p}-1}\right)\right\}$ of size $N_p$, that estimates the probability $\operatorname{Pr}[\mathbf{T}|Z=z]$. In deep learning, the goal of the training stage is to learn the neural network's trainable parameters $\boldsymbol{\theta}^{\prime}$, which minimize the empirical error represented by a loss function $\ell$:
\begin{equation}
\label{eq1}
\boldsymbol{\theta}^{\prime}=\underset{\boldsymbol{\theta}}{\operatorname{argmin}} \frac{1}{N_p} \sum_{i}^{N_p} \ell\left(F_{\boldsymbol{\theta}}\left(\mathbf{t}_{i}\right), z_{i}\right).
\end{equation}

Here, we regard the profiled SCA as a $c$-classification task, where $c$ denotes
the number of classes that depends on the leakage model. Additionally, we generally represent classes with one-hot encoding, i.e., each class is turned into a vector of $c$ values. Only one of $c$ values equals 1 and others equal 0, which denotes the membership of that class. The most common loss function in deep learning is the categorical cross-entropy (CCE):
\begin{equation}
\label{eq2}
L_{CCE}\left(F_{\boldsymbol{\theta}}, \mathcal{T}\right)=-\frac{1}{N_{p}} \sum_{i=1}^{N_{p}} \log _{2}\left(F_{\boldsymbol{\theta}}\left(\mathbf{t}_{i}\right)[z_i]\right),
\end{equation}
where the model $F_{\boldsymbol{\theta}}$ yields a probability vector for each input trace, consisting of the chances of each class to be selected, which is accomplished through the softmax layer.

During the attacking phase, the trained model would help predict the sensitive variables, hence the attacker can use an attack set of size $N_a$ to compute a score vector with the knowledge of the inputs for each key candidate. Actually, for each $k \in \mathcal{K}$, the log likelihood score is computed by
\begin{equation} 
\label{eq3}
s_{N_{a}}(k)=\log \left(\prod_{i=1}^{N_{a}} F\left(\mathbf{t}_{i}\right)[z_{i}]\right)=\sum_{i=1}^{N_{a}} \log \left(F\left(\mathbf{t}_{i}\right)[z_{i}]\right).
\end{equation}
Afterwards, the key hypothesis with the highest score is assumed to be the correct key. 

\subsection{Evaluation Metric}
\label{subsec2.3}

In the side channel community, success rate and key rank are two ordinarily used metrics for evaluating attack performance, and we adopt the latter in this paper. As illustrated in \eqref{eq3}, the attacker calculates a score for each key candidate, based on which he can sort all key candidates in decreasing order and acquire a key guess vector, denoted as $\mathbf{g}=\left(g_{1}, g_{2}, \ldots, g_{|\mathcal{K}|}\right)$. The key rank is defined as the position of $k^*$ in this vector, i.e.,
\begin{equation}
\label{eq4}
rank\left(k^{*}\right)=\sum_{k \in \mathcal{K}} \mathrm{1}_{s_{N_{a}}(k)>s_{N_{a}}\left(k^{*}\right)}
\end{equation}

In practice, we generally repeat the attack procedure certain times and take the average key rank as the evaluation metric. Naturally, we consider $g_{1}$ as the most possible key candidate and $g_{|\mathcal{K}|}$ as the least possible one. Therefore $g_{1}$ corresponds to $k^*$ indicates that key rank equals 0 and the attack is successful. Intuitively, we intend to achieve key rank equal to 0 with attack traces as few as possible.

\subsection{Datasets}
\label{subsec2.4}

This paper considers four publicly available datasets for subsequent experiments. In this section, we give some details about the employed datasets.

\subsubsection{ASCAD Dataset}
\label{subsubsec2.4.1} 
ASCAD\footnote{This dataset is available at \url{ https://github.com/ANSSI-FR/ASCAD}. \label{footnote1}} dataset consists of measurements collected from an 8-bit micro-controller running a software protected implementation of AES-128, where masking is applied as a countermeasure hence the security against first-order SCA is guaranteed. This dataset comprises 60000 traces, each of which involves 700 time samples. Among them, 50000 traces compose the profiling set and 10000 traces are in the attacking set \cite{Prouff2018StudyOD}. We intend to attack the third key byte and target the third Sbox operations in the first encryption round, then the label can be written as:
\begin{equation}
\label{eq11}
Y\left(k^{*}\right) = L(Sbox\left[P_{2} \oplus k^{*}\right]),
\end{equation}
where $L$ denotes the chosen leakage model, such as identity model.

\subsubsection{ASCAD Datasets with Desynchronization}
\label{subsubsec2.4.2} 
ASCAD\_desync50\textsuperscript{\ref{footnote1}} is composed of traces that are desynchronized from the original traces in ASCAD with a 50 samples maximum window. The trace desynchronization is simulated artificially by a python script called ASCAD\_generate.py\footnote{This script is available at \url{ https://github.com/ANSSI-FR/ASCAD}. \label{footnote2}}. In ASCAD\_generate.py. Similarly, ASCAD\_desync100\textsuperscript{\ref{footnote1}} contains measurements desynchronized with a 100 samples maximum window from the original traces. In these two datasets, there are still 50000 training traces and 10000 testing traces. For each trace, the label remains identical to the corresponding label in ASCAD dataset. Besides, we still attack the third key byte. 

\subsubsection{AES\_RD Dataset}
\label{subsubsec2.4.3} 
AES\_RD\footnote{At \url{https://github.com/ikizhvatov/randomdelays-traces}, this dataset is available. \label{footnote3}} dataset is a collection of traces measured from a protected software implementation of AES running over a smartcard of 8-bit Atmel AVR microcontroller. The adopted countermeasure is random delay \cite{Jeans2009An}, which induces trace misalignment and magnifies the handicap of an attack. This dataset possesses 50000 traces, from which we divide out 40000 traces for training and the other 10000 measurements for testing. There are 3500 features in each trace, hence the training procedure of CNN becomes much more time-consuming. For AES\_RD, we choose the first key byte to attack and  the first Sbox operation in the first AES round to compute the label:
\begin{equation}
\label{eq12}
Y\left(k^{*}\right) = L(Sbox\left[P_{0} \oplus k^{*}\right]).
\end{equation}

\section{Enhancing DL-based SCA with \textit{Mixup}}
\label{sec3}

In this section, we introduce the \textit{mixup} technique and narrate the general fashion of exploiting \textit{mixup} in DL-based SCA. Furthermore, to evaluate the quality of generated traces, we execute CPA analysis to decide whether the generated traces encompass meaningful information, i.e., leakage points.

\subsection{\textit{Mixup}}
\label{subsec3.1}
A supervised learning task proceeds with the intent of searching a function $F \in \mathcal{F}$ to characterize the relation between an input vector $\mathbf{X}$ and a label vector $\mathbf{Y}$, that satisfy the joint distribution $Pr(\mathbf{X}, \mathbf{Y})$. With the training dataset $\mathcal{D}=\left\{\left(\mathbf{x}_{i}, \mathbf{y}_{i}\right)\right\}_{i=1}^{n},$ where $\left(\mathbf{x}_{i}, \mathbf{y}_{i}\right) \sim Pr$, the empirical distribution $Pr$ may be estimated by
\begin{equation}
\label{eq5}
Pr_{\delta}(\mathbf{X}, \mathbf{Y})=\frac{1}{n} \sum_{i=1}^{n} \delta\left(\mathbf{X}=\mathbf{x}_{i}, \mathbf{Y}=\mathbf{y}_{i}\right)
\end{equation}
where $\delta\left(\mathbf{X}=\mathbf{x}_{i}, \mathbf{Y}=\mathbf{y}_{i}\right)$ is a Dirac mass centered at $\left(\mathbf{x}_{i}, \mathbf{y}_{i}\right)$.
Then we can approximate the expected risk as the empirical risk:
\begin{equation}
\label{eq6}
R_{\delta}(F)=\int \ell(F(\mathbf{X}), \mathbf{Y}) \mathrm{d} Pr_{\delta}(\mathbf{X}, \mathbf{Y})=\frac{1}{n} \sum_{i=1}^{n} \ell\left(F\left(\mathbf{x}_{i}\right), \mathbf{y}_{i}\right),
\end{equation}
where $\ell$ is the loss function. The Empirical Risk Minimization (ERM) principle refers to learning $f$ by minimizing \eqref{eq6} \cite{DBLP:journals/corr/abs-1710-09412}.

Notwithstanding the commonality of the ERM, there are still some alternatives, such as the Vicinal Risk Minimization (VRM) principle \cite{DBLP:journals/corr/abs-1710-09412}, in which the distribution $Pr$ is estimated by
\begin{equation}
\label{eq7}
Pr_{\nu}(\tilde{\mathbf{X}}, \tilde{\mathbf{Y}})=\frac{1}{n} \sum_{i=1}^{n} \nu\left(\tilde{\mathbf{x}}, \tilde{\mathbf{y}} \mid \mathbf{x}_{i}, \mathbf{y}_{i}\right),
\end{equation}
where $\nu$ is a vicinity distribution that quantifies the eventuality of finding the virtual input-label touple $(\tilde{\mathbf{X}}, \tilde{\mathbf{Y}})$ in the vicinity of the training pair $\left(\mathbf{x}_{i}, \mathbf{y}_{i}\right)$. Analogous to the ERM, the VRM minimizes the empirical vicinal risk instead with the generated dataset $\mathcal{D}_{\nu}:=\left\{\left(\tilde{\mathbf{x}}_{i}, \tilde{\mathbf{y}}_{i}\right)\right\}_{i=1}^{m}$:
\begin{equation}
\label{eq8}
R_{\nu}(F)=\frac{1}{m} \sum_{i=1}^{m} \ell\left(F\left(\tilde{\mathbf{x}}_{i}\right), \tilde{\mathbf{y}}_{i}\right).
\end{equation}
Particularly, \textit{mixup} is a generic vicinal distribution:
\begin{equation}
\label{eq9}
\begin{aligned} 
\mu(\tilde{x}, \tilde{y})=\frac{1}{n} \sum_{j}^{n} \mathbb{E}[\delta(\tilde{x}&=\lambda \cdot x_{i}+(1-\lambda) x_{j}, \\ &\left.\left.\tilde{y}=\lambda \cdot y_{i}+(1-\lambda) y_{j}\right)\right] 
\end{aligned}
\end{equation}
where $\lambda \sim \operatorname{Beta}(\alpha, \alpha),$ for $\alpha \in(0, \infty)$. Consequently, following this principle we can yield virtual input-label vectors
\begin{equation}
\label{eq10}
\begin{split}
\tilde{\mathbf{x}}=\lambda \mathbf{x}_{i}+(1-\lambda) \mathbf{x}_{j},
\\	
\tilde{\mathbf{y}}=\lambda \mathbf{y}_{i}+(1-\lambda) \mathbf{y}_{j},
\end{split}
\end{equation}
where $\left(\mathbf{x}_{i}, \mathbf{y}_{i}\right)$ and $\left(\mathbf{x}_{j}, \mathbf{y}_{j}\right)$ are two input-label pairs randomly picked up from training data, $\mathbf{y}_{i}$ and $\mathbf{y}_{j}$ take the form of one-hot encoding and $\lambda \in[0,1]$. The crucial parameter $\alpha$ regulates the intensity of interpolation between input-label vectors, reinstating the ERM when $\alpha \rightarrow 0$.

The \textit{mixup} principle is a pattern of data augmentation technique that facilitates the model $f$ to behave linearly amid training examples. And this linear practice is supposed to diminish the unsatisfactory prediction fluctuations as confronted with examples distributed differently from training data. Besides, reflecting on Occam's Razor, linearity appears to be a reasonable inductive bias on account of its universality as one of the simplest possible behaviors. An extensive evaluation was conducted in \cite{DBLP:journals/corr/abs-1710-09412} and the authors concluded that \textit{mixup} reduced the generalization error of state-of-the-art models on ImageNet, CIFAR, speech, and tabular datasets. Given the advantages of \textit{mixup} in both 1D and 2D data, we speculate that this technique may adapt to side channel measurements as well.  

\subsection{Data Augmentation with \textit{Mixup}}
\label{subsec3.2}

\begin{figure*}[htbp]
	\centering   
	\includegraphics[width=0.9\textwidth]{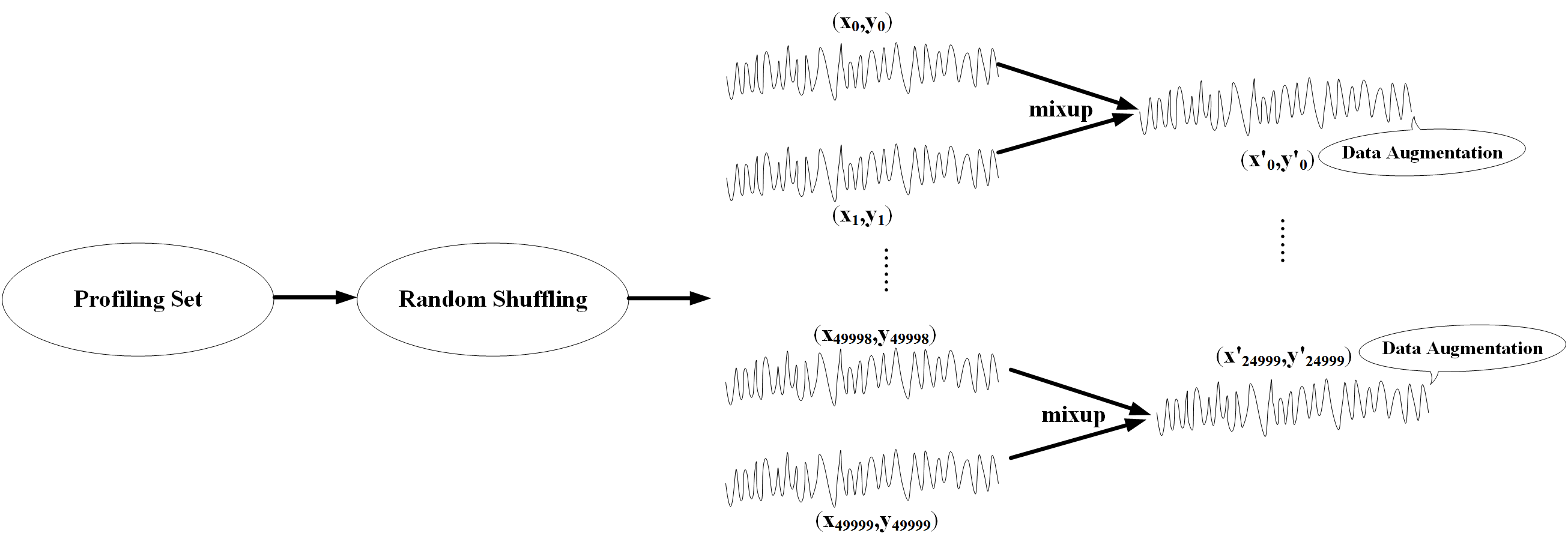} 
	\caption{The General Design of Enhancing DL-based SCA with \textit{Mixup}.}
	\label{fig17}
\end{figure*}

The general design of enhancing DL-based SCA with \textit{mixup} is illustrated in Fig. \ref{fig17}. We have always been underlining that our jumping-off point is to mitigate the issues caused by restricted measurements. From this perspective, we divide out merely a portion of profiling traces to simulate the constrained scenario. For each dataset, we respectively pick up 3000, 5000, 10000 and all measurements from the training set at random, to compare the performance of original dataset and augmented dataset in distinct settings, which we will expatiate in section \ref{sec4}. Afterwards, we perform \textit{mixup} to the selected traces. Actually, we have considered combining more than two input-label touples, which means the training pair are generated according to
\begin{center}
	$\tilde{\mathbf{x}}=\lambda_1 \mathbf{x}_{1}+\lambda_2 \mathbf{x}_{2}+...+(1-\sum\limits_{i=1}^{k-1}) \mathbf{x}_{k}$,
	
	$\tilde{\mathbf{y}}=\lambda_1 \mathbf{y}_{1}+\lambda_2 \mathbf{y}_{2}+...+(1-\sum\limits_{i=1}^{k-1}) \mathbf{y}_{k}$,
\end{center}
where $\left(\mathbf{x}_{i}, \mathbf{y}_{i}\right), i=1,...,k$, are $k$ random input-label pairs in the profiling set, $\mathbf{x}_{i}$ denotes a power measurement in SCA context and $\lambda_{i} \in[0,1]$ for $i=1,...,k-1$. However, \cite{DBLP:journals/corr/abs-1710-09412} pointed out that convex combinations of three (or more) samples with weights following Dirichlet distribution, i.e., multivariate Beta distribution, does not earn further gain, but increases the computation cost of \textit{mixup} on the contrary. In this way, we still joint input-label vectors pairwise in light of \eqref{eq10}, and then merge initial and generated input-label pairs, which engenders an expanded dataset half larger than the original training set. As for the variable $\alpha$, we first set $\alpha$ to 0.5. Afterwards, we test 0.3 and 0.7 in some situations to complete the experiments. 

After obtaining an enlarged training set by implementing \textit{mixup}, we train two CNNs with original dataset and new \textit{mixup} dataset, respectively, in the premise of unchanged network architecture  and hyperparameters. Once the training process is accomplished, for each neural network, we select the finest epoch where model capability attains the best. Hereafter, still for each selected best model, the test procedure is repeated 50 times and the average key rank is calculated to evaluate attack performance, thereupon allowing us to contrast the power of original and new profiling set.  

Similar to the applications of \textit{mixup} in speech and image datasets, \textit{mixup} can generate virtual side channel measurements linearly. In this way, the generalization error is supposed to be decreased. On the other hand, the profiling set is augmented, which encourages to train the network comprehensively especially in restricted situations, such as the circumstance where there are merely 3000 profiling traces.

\subsection{Analysis of Generated Datasets}
\label{subsec3.3}

\begin{figure}[tb]
	\centering	
	\subfloat[Original training set.]
	{
		\begin{minipage}[t]{0.25\textwidth}
			\centering   
			\includegraphics[width=1\textwidth]{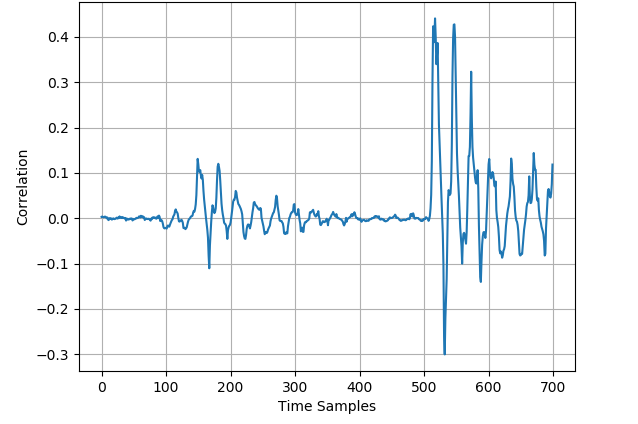} 
		\end{minipage}
	}
	\subfloat[Generated training set.]
	{
		\begin{minipage}[t]{0.25\textwidth}
			\centering 
			\includegraphics[width=1\textwidth]{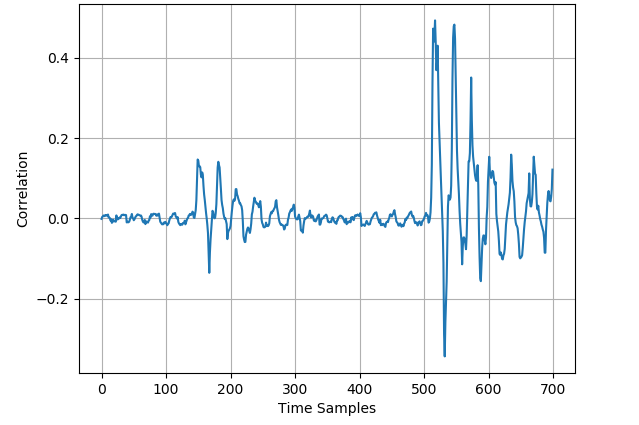} 
		\end{minipage}
	}	
	\caption{CPA Analysis for ASCAD.}
	\label{fig1}
\end{figure}

As stated in section \ref{subsec3.2}, we exploit \textit{mixup} to generate virtual traces. Naturally, there is a demand to explore whether the generated traces are appropriate. With regards to this, we carry out CPA analysis of initial and generated training set for several public datasets. Fig. \ref{fig1} illustrates the CPA analysis of the 50000 original profiling traces and the 25000 generated profiling traces for ASCAD dataset. Intuitively, these two CPA results are exceedingly indistinguishable in terms of the leakage location and the correlation intensity. Meticulously observing the figure, we notice that for the generated dataset the correlation at the highest leakage position is even somewhat higher, which signifies that the knowledge involved in generated traces is consistent with that of original traces. In view of CPA analysis, it seems considerably sensible to fulfill the anticipation of mounting a more efficient attack by employing \textit{mixup} to augment the profiling set.

\begin{figure}[tb]
	\centering	
	\subfloat[Original training set.]
	{
		\begin{minipage}[t]{0.25\textwidth}
			\centering   
			\includegraphics[width=1\textwidth]{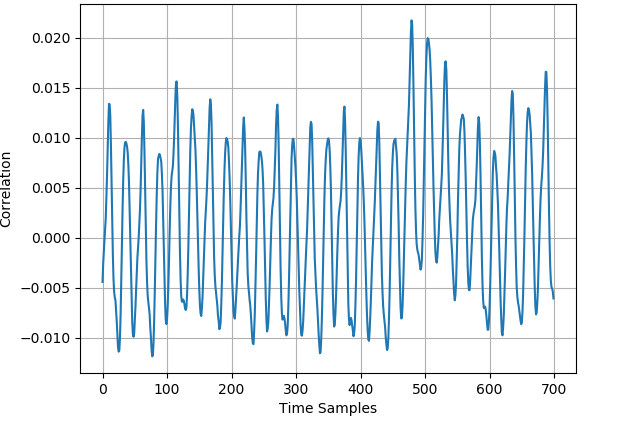} 
		\end{minipage}
	}
	\subfloat[Generated training set.]
	{
		\begin{minipage}[t]{0.25\textwidth}
			\centering 
			\includegraphics[width=1\textwidth]{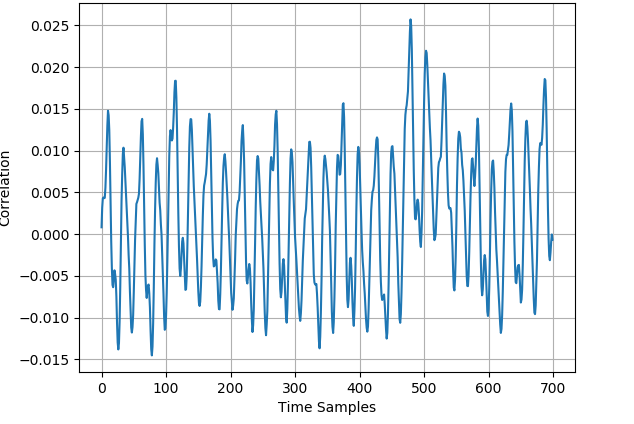} 
		\end{minipage}
	}	
	\caption{CPA Analysis for ASCAD\_desync50.}
	\label{fig2}
\end{figure}

\begin{figure}[tb]
	\centering	
	\subfloat[Original training set.]
	{
		\begin{minipage}[t]{0.25\textwidth}
			\centering   
			\includegraphics[width=1\textwidth]{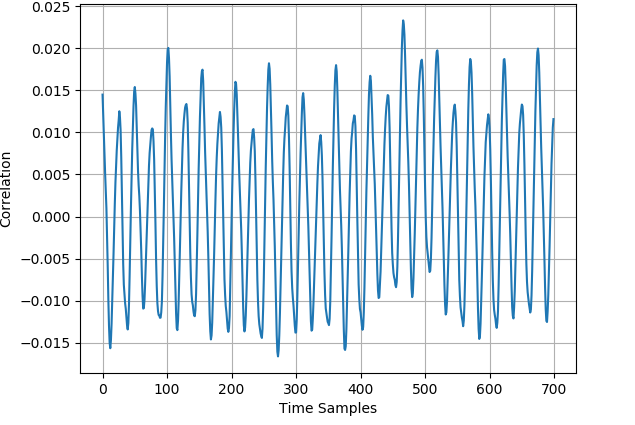} 
		\end{minipage}
	}
	\subfloat[Generated training set.]
	{
		\begin{minipage}[t]{0.25\textwidth}
			\centering 
			\includegraphics[width=1\textwidth]{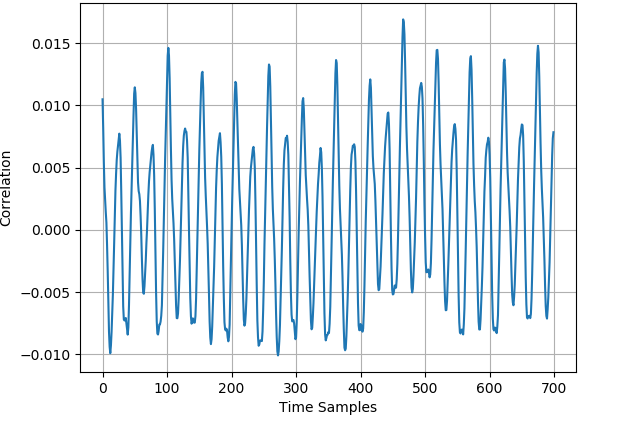} 
		\end{minipage}
	}	
	\caption{CPA Analysis for ASCAD\_desync100.}
	\label{fig3}
\end{figure}

\begin{figure}[tb]
	\centering	
	\subfloat[Original training set.]
	{
		\begin{minipage}[t]{0.25\textwidth}
			\centering   
			\includegraphics[width=1\textwidth]{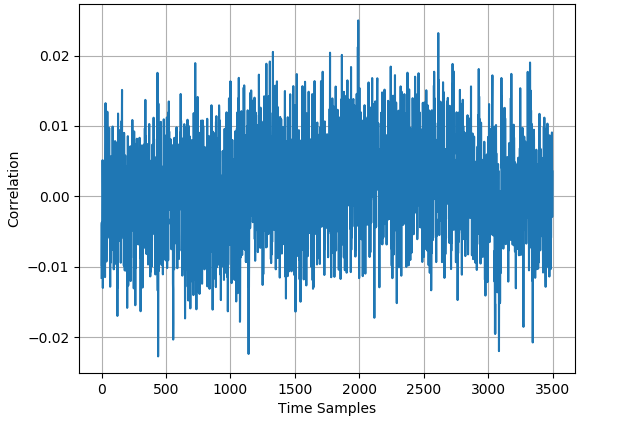} 
		\end{minipage}
	}
	\subfloat[Generated training set.]
	{
		\begin{minipage}[t]{0.25\textwidth}
			\centering 
			\includegraphics[width=1\textwidth]{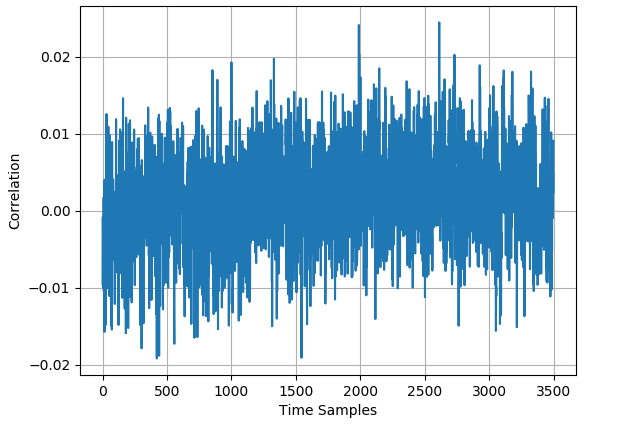} 
		\end{minipage}
	}	
	\caption{CPA Analysis for AES\_RD.}
	\label{fig4}
\end{figure}

The CPA analyses of the 50000 original profiling traces and the 25000 generated profiling traces for ASCAD\_desync50 dataset and ASCAD\_desync100 dataset are respectively displayed in Fig. \ref{fig2} and Fig. \ref{fig3}. For AES\_RD dataset, Fig. \ref{fig4} demonstrates the CPA results of 40000 original traces and the 20000 generated traces. In all three datasets, there exists trace misalignment, which can be recognized from the diffusely distributed spires in the figures. Notwithstanding, at peaks with different strengths, the CPA results of original dataset and generated dataset are corresponding, which exhibits the justifiability of \textit{mixup} once again.

\section{Exploiting \textit{Mixup} in Practice}
\label{sec4}

In this section, we apply \textit{mixup} technique to several  publicly available datasets. We test different leakage models and different values of \textit{mixup} parameter $\alpha$. Our experiments are running with the NVIDIA Graphics Processing Units (GPUs) and the open-source deep-learning library Keras and TensorFlow backend.

\subsection{The Selection of Leakage Model}
\label{subsec4.1}
\begin{figure*}[tb]
	\centering	
	\subfloat[ID model.]
	{
		\begin{minipage}[t]{0.33\textwidth}
			\centering   
			\includegraphics[width=0.9\textwidth]{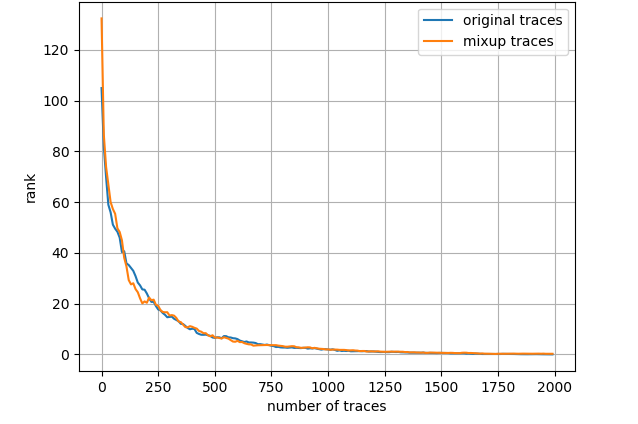} 
		\end{minipage}
	}
	\subfloat[HW model.]
	{
		\begin{minipage}[t]{0.33\textwidth}
			\centering 
			\includegraphics[width=0.9\textwidth]{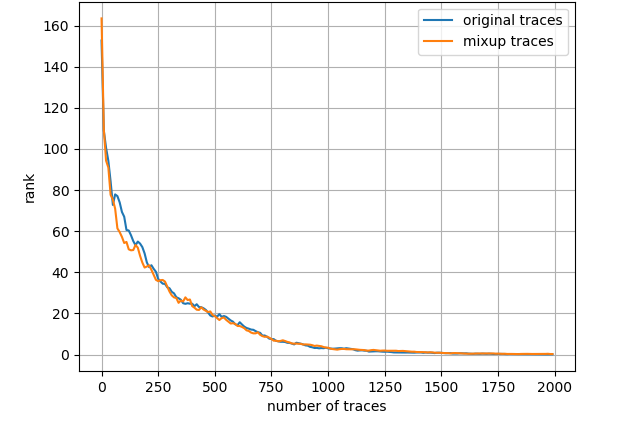} 
		\end{minipage}
	}
	\subfloat[LSB model.]
	{
		\begin{minipage}[t]{0.33\textwidth}
			\centering 
			\includegraphics[width=0.9\textwidth]{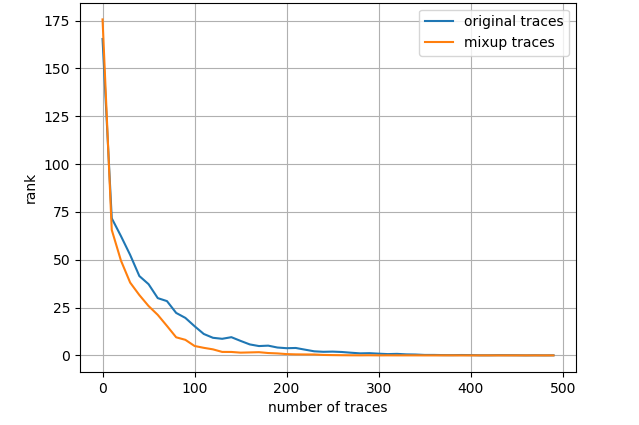} 
		\end{minipage}
	}
	\caption{Average key rank for ASCAD with different leakage models.}
	\label{fig5}
\end{figure*}

General leakage models include least significant bit (LSB) model, hamming weight (HW) model, identity (ID) model, etc. LSB model supposes the least significant bit of the targeted Sbox output to be the label, HW model calculates the hamming weight of the intermediate value while ID model straight considers the sensitive value itself as the label. We conduct tentative experiments with these leakage models and surprisingly discover that for all datasets LSB model performs immensely superiorly, which is slightly counter-intuitive since this model is practically left out in related works. Fig. \ref{fig5} shows the average key rank results for ASCAD dataset when different leakage models are employed, where all of 50000 profiling traces are fed into the network and the \textit{mixup} parameter $\alpha$ equals 0.5. In each figure, the orange line represents the attacking results for \textit{mixup} training set whereas the blue line stands for the original profiling set. With respect to the network architecture, we exclusively adopt the CNN proposed in \cite{Prouff2018StudyOD}, an uncomplicated CNN designed from VGG-16 fashion, that comprises five convolution layers, five following pooling layers and three fully-connected layers, the final one of which is the softmax layer aiming at figuring out the probabilities for each class. Other hyperparameters, including loss function, optimizer, learning rate, are correspondingly specified as categorical cross entropy, RMSprop and 0.00001. For all experiments in this paper, the CNN structure and hyperparameters stand unaltered apart from the epoch, for which we select the best one to cease training in every scenario, as claimed in section \ref{subsec3.2}. The selected finest epochs in this section can be consulted from Appendix \ref{Appendix}. 

From Fig. \ref{fig5} it is investigated that the attack performance is evidently better with LSB model. Approximately 1250 traces and 1500 traces are required to achieve average key rank 0 for ID model and HW model, respectively, however, roughly 200 traces are adequate to reach the same condition for LSB model when \textit{mixup} is adopted. On the other hand, it appears that when using ID model and HW model, \textit{mixup} does not bring about any performance gain. We suppose that this is because 50000 profiling traces are already fairly sufficient and ASCAD dataset is not a particularly thorny dataset to DL\_based SCA. Subsequent experimental results demonstrate that when the dataset is tougher to handle or the amount of profiling traces becomes restricted, \textit{mixup} will play a more crucial role in enhancing attack performance. Note that this section is especially emphasizing the ascendancy of LSB model. 

\begin{figure*}[tb]
	\centering	
	\subfloat[ID model.]
	{
		\begin{minipage}[t]{0.33\textwidth}
			\centering   
			\includegraphics[width=0.9\textwidth]{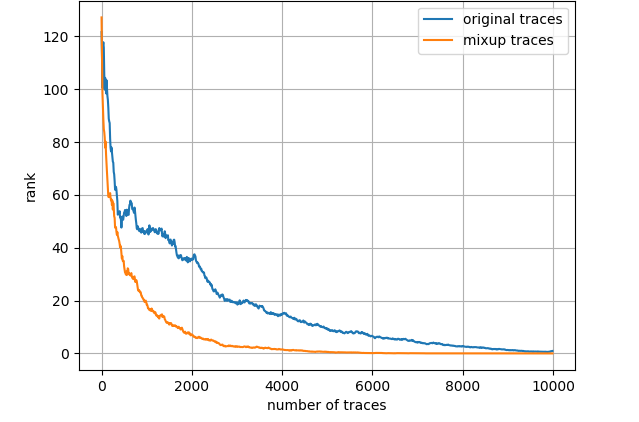} 
		\end{minipage}
	}
	\subfloat[HW model.]
	{
		\begin{minipage}[t]{0.33\textwidth}
			\centering 
			\includegraphics[width=0.9\textwidth]{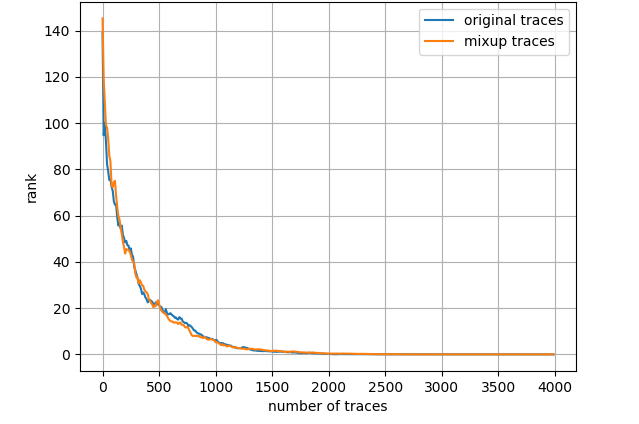} 
		\end{minipage}
	}
	\subfloat[LSB model.]
	{
		\begin{minipage}[t]{0.33\textwidth}
			\centering 
			\includegraphics[width=0.9\textwidth]{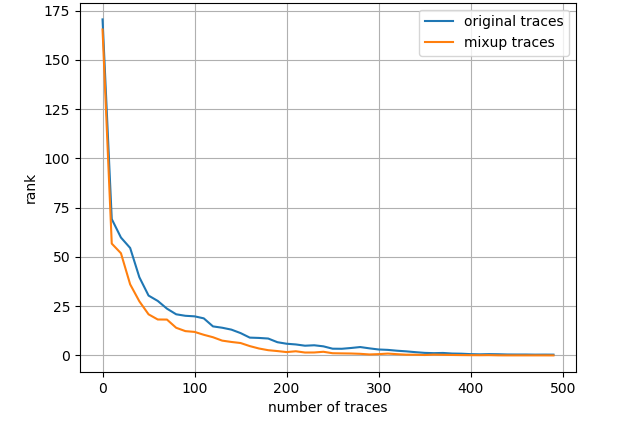} 
		\end{minipage}
	}
	\caption{Average key rank for ASCAD\_desync50 with different leakage models.}
	\label{fig6}
\end{figure*}

\begin{figure*}[tb]
	\centering	
	\subfloat[ID model.]
	{
		\begin{minipage}[t]{0.33\textwidth}
			\centering   
			\includegraphics[width=0.9\textwidth]{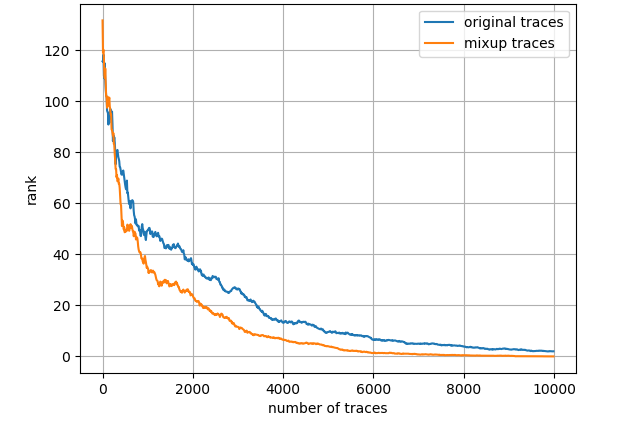} 
		\end{minipage}
	}
	\subfloat[HW model.]
	{
		\begin{minipage}[t]{0.33\textwidth}
			\centering 
			\includegraphics[width=0.9\textwidth]{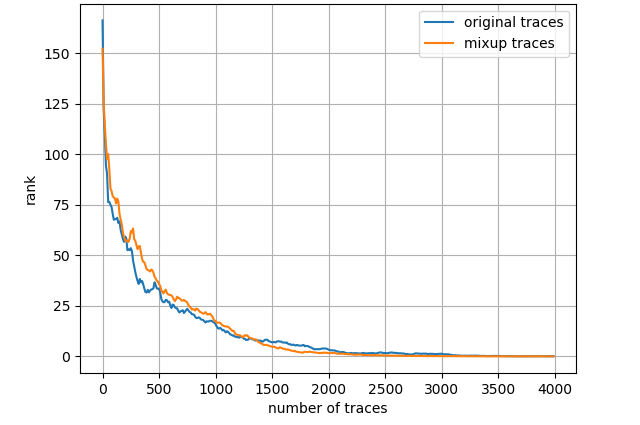} 
		\end{minipage}
	}
	\subfloat[LSB model.]
	{
		\begin{minipage}[t]{0.33\textwidth}
			\centering 
			\includegraphics[width=0.9\textwidth]{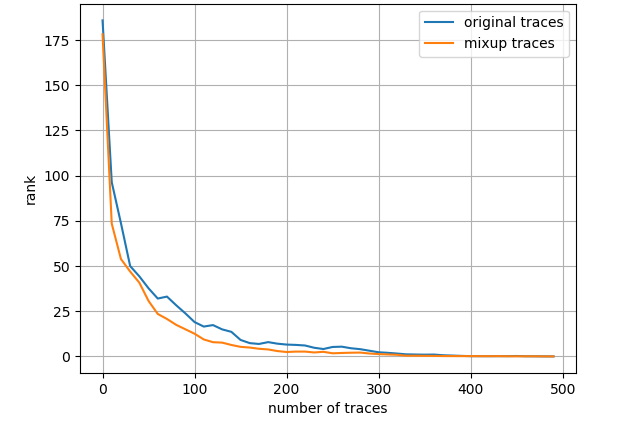} 
		\end{minipage}
	}
	\caption{Average key rank for ASCAD\_desync100 with different leakage models.}
	\label{fig7}
\end{figure*}

\begin{figure*}[tb]
	\centering	
	\subfloat[ID model.]
	{
		\begin{minipage}[t]{0.33\textwidth}
			\centering   
			\includegraphics[width=0.9\textwidth]{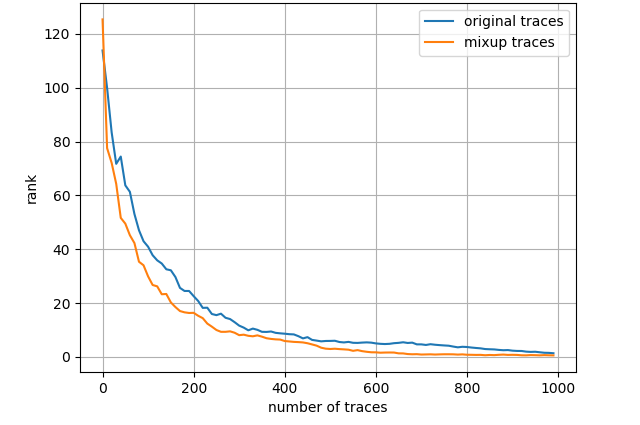} 
		\end{minipage}
	}
	\subfloat[HW model.]
	{
		\begin{minipage}[t]{0.33\textwidth}
			\centering 
			\includegraphics[width=0.9\textwidth]{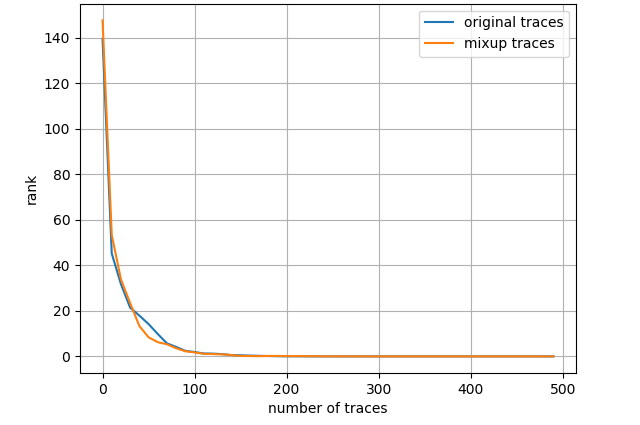} 
		\end{minipage}
	}
	\subfloat[LSB model.]
	{
		\begin{minipage}[t]{0.33\textwidth}
			\centering 
			\includegraphics[width=0.9\textwidth]{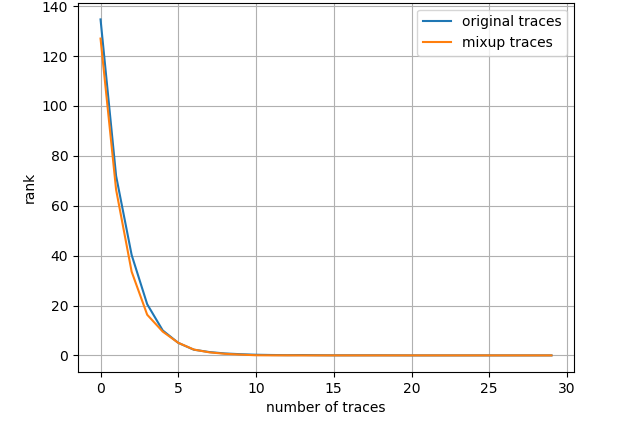} 
		\end{minipage}
	}
	\caption{Average key rank for AES\_RD with different leakage models.}
	\label{fig8}
\end{figure*}

Fig. \ref{fig6} depicts the attack results for ASCAD\_desync50 dataset and three leakage models, also with the entire training set. Regarding ID model, \textit{mixup} boosts the results to a great extent then about 4000 measurements are needed to launch a successful SCA. As to HW model, \textit{mixup} does not promote the results apparently and we demand less than 2000 traces in both scenarios. Nevertheless, turning to LSB model, merely less than 300 traces are enough to realize an average rank of 0 for the network trained with \textit{mixup} dataset. The experimental results for ASCAD\_desync100 dataset are depicted in Fig. \ref{fig7}, which is similar to Fig. \ref{fig6}. Under three leakage models, \textit{mixup} yields performance improvement in some degree and the required amount of traces are correspondingly about 7000, 2200 and 300 for ID, HW and LSB models. AES\_RD is the easiest dataset, which is implied by the average rank results depicted in Fig. \ref{fig8}. This dataset is considerably simple so \textit{mixup} is barely influential for ID model since all 40000 traces are utilized to train. In addition, to break AES\_RD, 800, 150 and 8 traces are needed under the distinct leakage models. All these results explicitly indicate that LSB has a noticeable advantage among these leakage models. We have not figured out the exact justification to account for this phenomenon. But we can absorb some inspirations from \cite{9006657}, which proposed a multi-label classification model, essentially the naive ensemble pattern of eight single-bit models, and acquired remarkable results for ASCAD even in high-desynchronization situations. We conjecture that the multi-label model excellence is partially associated with LSB as least significant bit model transcends all other seven single-bit models in their experiments \cite{9006657}. Besides, our average rank results are approaching theirs although we just apply LSB. On the whole, we decide LSB model to be the option in the following experiments. 

\subsection{ASCAD Dataset}
\label{subsec4.2}
According to the general design described in section \ref{subsec3.2}, we use 3000, 5000, 10000 and all profiling traces to perform \textit{mixup}, where the parameter $\alpha$ is set to 0.5. Table \ref{table1} manifests the average rank results for ASCAD when different numbers of original and \textit{mixup} profiling traces are used. The determined epochs to early-stop are also tabulated in Appendix \ref{Appendix}. Besides, $\alpha$ in sections \ref{subsec4.3} - \ref{subsec4.4} keeps unchanged and the corresponding epochs for experiments in sections \ref{subsec4.3} - \ref{subsec4.4} are also displayed in Appendix \ref{Appendix}. 

From Table \ref{table1} it is observed that the performance gain is more apparent for inadequate profiling traces, which is reflected in the gap of 540 and the gap of 300 attacking traces for 3000 and 5000 original profiling traces, respectively. When the size of training set increases, there appears a decline in the performance gap, nonetheless the \textit{mixup} training set still outpaces the original one to some extent. This is readily comprehensible since to train a neural network with multiple parameters, only with sufficient training data can the network extract appropriate knowledge from features to map a never seen input vector to a proper label vector. Furthermore, section \ref{subsec3.3} has revealed the consistency of \textit{mixup} traces and original traces in terms of leakage information. Consequently, in restricted scenarios, \textit{mixup} dataset can exactly relieve an insufficient predicament, where underfitting happens, thus encouraging a successful attack. On the other hand, in comparatively sufficient situations, \textit{mixup} dataset further favors comprehensive learning on the existing basis and yields modest performance enhancement.

\begin{table}[tb]
	\centering
	\caption{Average key rank results of ASCAD for different number of original and \textit{mixup} profiling traces.}
	\label{table1}
	\normalsize
	\tabcolsep 3pt 
	\renewcommand{\arraystretch}{1.3} 
	\begin{tabular*}{7.5cm}{c|p{1cm}<{\centering}|p{1cm}<{\centering}|p{1cm}<{\centering}|p{1cm}<{\centering}}
		\toprule[1.5pt]
		& 3000 & 5000 & 10000 & 50000
		\\[2pt]  \hline
		Original Dataset & 1402 & 702 & 327 & 299 \\  \hline
		\textit{Mixup} Dataset & \textcolor{red}{862} & \textcolor{red}{402} & \textcolor{red}{290} & \textcolor{red}{192} \\ 
		\bottomrule[1.5pt]
	\end{tabular*}	
\end{table}

\subsection{ASCAD Datasets with Desynchronization}
\label{subsec4.3}

Next, we shift the attention to ASCAD datasets with desynchronization. In Table \ref{table2}, we depict the average rank results for ASCAD\_desync50 when a different number of original and \textit{mixup} profiling traces are used, which is somewhat similar to the results in section \ref{subsec4.2}. For 3000, 5000, 10000 and 50000 original training traces, the reduction of required attacking traces induced by \textit{mixup} is correspondingly 357, 381, 143 and 100. We suppose this is perhaps because CNNs is advantageous in processing asynchronous data and desynchronization of 50 is still quite within CNNs' reach. 

When it turns to ASCAD\_desync100 dataset, Table \ref{table3} display our experimental results. In such a high-desynchronization scenario, successfully mounting SCA is notably harder, especially when the size of profiling set is small. It is shown that if barely 3000 original profiling traces are available, 6279 attacking traces are needed to achieve an average key rank equal to 0, whereas with \textit{mixup} 3453 measurements are already adequate to break ASCAD\_desync100, which means the attack performance has approximately doubled. Likewise, for 5000 original profiling traces, the size of required attacking set drops from 3618 to 1753 owing to \textit{mixup}. On account of the high desynchronization, unlike the aforementioned two datasets, now 10000 profiling traces are not quite sufficient and \textit{mixup} brings about an improvement of 654 attacking traces. In light of these results, we deduce that especially for troublesome conditions, i.e., inadequate profiling traces or difficult datasets, \textit{mixup} is more capable of exerting the advantages. 

\begin{table}[tb]
	\centering
	\caption{Average key rank results of ASCAD\_desync50 for different number of original and \textit{mixup} profiling traces.}
	\label{table2}
	\normalsize
	\tabcolsep 3pt 
	\renewcommand{\arraystretch}{1.3} 
	\begin{tabular*}{7.5cm}{c|p{1cm}<{\centering}|p{1cm}<{\centering}|p{1cm}<{\centering}|p{1cm}<{\centering}}
		\toprule[1.5pt]
		& 3000 & 5000 & 10000 & 50000
		\\[2pt]  \hline
		Original Dataset & 1155 & 1163 & 666 & 381 \\ \hline
		\textit{Mixup} Dataset & \textcolor{red}{798} & \textcolor{red}{782} & \textcolor{red}{523} & \textcolor{red}{281} \\  
		\bottomrule[1.5pt]
	\end{tabular*}	
\end{table}

\begin{table}[tb]
	\centering
	\caption{Average key rank results of ASCAD\_desync100 for different number of original and \textit{mixup} profiling traces.}
	\label{table3}
	\normalsize
	\tabcolsep 3pt 
	\renewcommand{\arraystretch}{1.3} 
	\begin{tabular*}{7.5cm}{c|p{1cm}<{\centering}|p{1cm}<{\centering}|p{1cm}<{\centering}|p{1cm}<{\centering}}
		\toprule[1.5pt]
		& 3000 & 5000 & 10000 & 50000
		\\[2pt]  \hline
		Original Dataset & 6279 & 3618 & 1342 & 362 \\  \hline
		\textit{Mixup} Dataset & \textcolor{red}{3453} & \textcolor{red}{1753} & \textcolor{red}{688} & \textcolor{red}{320} \\
		\bottomrule[1.5pt]
	\end{tabular*}	
\end{table}

\subsection{AES\_RD Dataset}
\label{subsec4.4}

Finally, we illustrate the average rank results for AES\_RD in Table \ref{table4}. Despite the existence of desynchronization, this dataset is fairly easy, which is clearly implied by the small quantity of required attacking traces in all settings. Even though there are merely 3000 original profiling traces for training the network, 35 attacking traces can already break AES\_RD, yet \textit{mixup} facilitates a more efficient attack, where only 19 measurements are enough. When the entire training set is used, respectively 10 and 8 traces can reach an average rank of 0 for original dataset and \textit{mixup} dataset. In another two settings, \textit{mixup} produces some gain of a few attacking traces as well, which again confirms that \textit{mixup} is favorable for enhancing attack performance.

\begin{table}[t]
	\centering
	\caption{Average key rank results of AES\_RD for different number of original and \textit{mixup} profiling traces.}
	\label{table4}
	\normalsize
	\tabcolsep 3pt 
	\renewcommand{\arraystretch}{1.3} 
	\begin{tabular*}{7.5cm}{c|p{1cm}<{\centering}|p{1cm}<{\centering}|p{1cm}<{\centering}|p{1cm}<{\centering}}
		\toprule[1.5pt]
		& 3000 & 5000 & 10000 & 50000
		\\[2pt]  \hline
		Original Dataset & 35 & 20 & 14 & 10 \\  \hline
		\textit{Mixup} Dataset & \textcolor{red}{19} & \textcolor{red}{16} & \textcolor{red}{10} & \textcolor{red}{8} \\
		\bottomrule[1.5pt]
	\end{tabular*}	
\end{table}

\subsection{Impact of \texorpdfstring{$\alpha$}{}}
\label{subsec4.5}

\begin{table*}[tb]
	\centering
	\caption{Average key rank results of four datasets for different number of original and \textit{mixup} profiling traces, and different values of $\alpha$.}
	\label{table5}
	\small
	\tabcolsep 3pt 
	\renewcommand{\arraystretch}{1.3} 	
	\begin{tabular}{c|p{1.3cm}<{\centering}|p{1.3cm}<{\centering}|p{1.3cm}<{\centering}|p{1.3cm}<{\centering}|p{1.3cm}<{\centering}|p{1.3cm}<{\centering}|p{1.3cm}<{\centering}|p{1.3cm}<{\centering}}
		\toprule[1.5pt]
		\multirow{2}{*}{\diagbox[width=6.5em,trim=l]{$\alpha$}{Dataset}} & \multicolumn{2}{p{2.6cm}<{\centering}|}{ASCAD} & \multicolumn{2}{p{2.6cm}<{\centering}|}{ASCAD\_desync50} & \multicolumn{2}{p{2.6cm}<{\centering}|}{ASCAD\_desync100} & \multicolumn{2}{p{2.6cm}<{\centering}}{AES\_RD}
		\\[2pt]  \cline{2-9}  
		& 3000 & 50000 & 3000 & 50000 & 3000 & 50000 & 3000 & 50000 
		\\[2pt]  \hline
		$\alpha=0.3$ & \textcolor{red}{815} & \textcolor{red}{157} & 1183 & \textcolor{red}{233} & 3358 & \textcolor{red}{252} & 24 & 9 \\  \hline
		$\alpha=0.7$ & 931 & 200 & 1612 & 246 & \textcolor{red}{1606} & 361 & 23 & 9 \\  \hline
		$\alpha=0.5$ & 862 & 192 & \textcolor{red}{798} & 281 & 3453 & 320 & \textcolor{red}{19} & \textcolor{red}{8} \\  \hline
		Original Dataset & 1402 & 299 & 1155 & 381 & 6279 & 362 & 35 & 10 \\  
		\bottomrule[1.5pt]
	\end{tabular}
\end{table*}

For experiments in sections \ref{subsec4.1} - \ref{subsec4.4}, we always set $\alpha$ to 0.5. To investigate the impact of $\alpha$ in \textit{mixup}, we also try out other values including 0.3 and 0.7. In this section, we just select 3000 and 50000 traces from original training set at random to perform \textit{mixup}. In Table \ref{table5}, we comprehensively compare key rank results for four datasets when $\alpha$ is set to 0.3, 0.5, 0.7 and \textit{mixup} is not exploited. 

For ASCAD dataset, we find that when $\alpha$ equals to 0.3, the brought performance gain even slightly outstrips the case where $\alpha$ is setting to 0.5. Specifically, for $\alpha$ equal to 0.3, when there are 3000 original profiling traces, \textit{mixup} reduces the amount of attacking traces from 1402 to 815; when all profiling traces are utilized, \textit{mixup} brings down the number from 299 to 157. While the performance gain engendered by $\alpha$ equal to 0.5 in these two settings are separately 540 and 107, as stated in section \ref{subsec4.2}. On the other hand, $\alpha$ set to 0.7 performs a bit poorer than 0.5, where the attacking trace decreasements are 431 and 99 correspondingly in two cases. 

For ASCAD\_desync50 dataset, when using 3000 original profiling traces, the results of \textit{mixup} dataset are quite weird since we even demand more attacking traces compared to original profiling set and this phenomenon is more serious for $\alpha$ equal to 0.7. As for the 50000 profiling trace case, it turns a bit normal as $\alpha$ equal to 0.3 and 0.7 respectively yields performance increase of 148 and 135 measurements, both moderately superior to the increase of 100 measurements produced by $\alpha$ equal to 0.5.

For ASCAD\_desync100 dataset, when $\alpha$ is set to 0.3, the performance enhancement is similar to the 0.5 case. While for $\alpha$ set to 0.7, if 3000 original traces are available, the enhancement is surprisingly remarkable, which means a reduction of measurements from 6279 to 1606, much more powerful than a gap of 2826 measurements in 0.5 case. If we use 50000 original traces, the enhancement becomes neglectable contrarily. For AES\_RD dataset, three values of $\alpha$ perform analogously, where $\alpha$ equal to 0.5 somewhat exceeds another two settings. Concluding the results we claim that in general all these three $\alpha$ are able to favor more efficient SCA and $\alpha$ equal to 0.5 performs most steadily despite $\alpha$ set to 0.3 and 0.7 can bring about more significant performance gain in some circumstances. Therefore, we still recommend 0.5 for \textit{mixup} parameter. Certainly, how other values of $\alpha$ influence attack performance remains to be explored, which we may investigate in the future.

\section{Conclusions}
\label{sec5}

In this paper, we discuss a DA technique---\textit{mixup}, which is initially designed to mitigate the difficulties in a restricted position for large deep neural networks. Considering the practical SCA scenario, where profiling traces are probably insufficient due to some restrictions such as time and resource limits, we propose to exploit \textit{mixup} in DL-based SCA context, for the sake of enlarging training set thus increasing the chances of mounting a successful attack. We conduct experiments on four public datasets, test different values of \textit{mixup} parameter $\alpha$, and verify that our proposition is feasible. Generally, executing \textit{mixup} with all three values is capable of improving attack performance by reducing the required attacking traces, among which $\alpha$ equal to 0.5 performs the most steadily. Besides, we find that LSB model is much more superior to prevalently used ID model and HW model, which is never emphasized in previous works.

In the future, we plan to extend our analysis to more datasets and explore whether there is a regular pattern revealing how values of parameter $\alpha$ affect attack performance. What's more, we want to figure out the reason why LSB model possesses such counter-intuitive performance even though this model simply utilizes a single bit of the sensitive variable.

\bibliographystyle{IEEEtran}
\bibliography{IEEEabrv,mybibfile}

\appendix
\section{}
\label{Appendix}

To save space, we use desync50, desync50, RD, Orig, \textit{Mixup} to represent ASCAD\_desync50, ASCAD\_desync100, AES\_RD, original and \textit{mixup} dataset, respectively.
\begin{table}[hb]
	\centering
	\caption{The selected best epochs for experiments in section \ref{subsec4.1}.}
	\label{table6}
	\small
	\tabcolsep 1.5pt 
	\renewcommand{\arraystretch}{1.3} 	
	\begin{tabular}{c|c|c|c|c|c|c|c|c}
	   \toprule[1.5pt]
		\multirow{2}{*}{\diagbox[width=6em,trim=l]{Model}{Dataset}} & \multicolumn{2}{c|}{ASCAD} & \multicolumn{2}{c|}{desync50} & \multicolumn{2}{c|}{desync100} & \multicolumn{2}{c}{RD}
		\\[2pt]  \cline{2-9}  
		& Orig & \textit{Mixup} & Orig & \textit{Mixup} & Orig & \textit{Mixup} & Orig & \textit{Mixup}
		\\[2pt]  \hline
		ID & 19 & 9 & 156 & 58 & 160 & 118 & 29 & 14 \\ \hline
		HW & 34 & 24 & 97 & 53 & 104 & 80 & 22 & 19 \\  \hline
		LSB & 21 & 8 & 30 & 24 & 39 & 33 & 15 & 12 \\  
	   \bottomrule[1.5pt]
	\end{tabular}
\end{table}	
\begin{table}[hb]
	\centering
	\caption{The selected best epochs for experiments in sections \ref{subsec4.2}-\ref{subsec4.4}.}
	\label{table7}
	\small
	\tabcolsep 1.5pt 
	\renewcommand{\arraystretch}{1.3} 	
	\begin{tabular}{c|c|c|c|c|c|c|c|c}
		\toprule[1.5pt]
		\multirow{2}{*}{\diagbox[width=6em,trim=l]{Orig}{Dataset}} & \multicolumn{2}{c|}{ASCAD} & \multicolumn{2}{c|}{desync50} & \multicolumn{2}{c|}{desync100} & \multicolumn{2}{c}{RD}
		\\[2pt]  \cline{2-9}  
		& Orig & \textit{Mixup} & Orig & \textit{Mixup} & Orig & \textit{Mixup} & Orig & \textit{Mixup}
		\\[2pt]  \hline
		3000 & 41 & 30 & 138 & 103 & 576 & 557 & 47 & 50 \\  \hline
		5000 & 47 & 26 & 154 & 100 & 236 & 238 & 32 & 27 \\  \hline
		10000 & 28 & 20 & 103 & 75 & 109 & 106 & 15 & 12 \\  \hline
		50000 & 21 & 8 & 30 & 24 & 39 & 33 & 15 & 12 \\
		\bottomrule[1.5pt]
	\end{tabular}
\end{table}
\begin{table}[hb]
	\centering
	\caption{The selected best epochs for experiments in section \ref{subsec4.5}.}
	\label{table8}
	\small
	\tabcolsep 1.5pt 
	\renewcommand{\arraystretch}{1.3} 	
	\begin{tabular}{c|p{0.8cm}<{\centering}|p{0.8cm}<{\centering}|p{0.8cm}<{\centering}|p{0.8cm}<{\centering}|p{0.8cm}<{\centering}|p{0.8cm}<{\centering}|p{0.8cm}<{\centering}|p{0.8cm}<{\centering}}
		\toprule[1.5pt]
		\multirow{2}{*}{\diagbox[width=6em,trim=l]{Orig}{Dataset}} & \multicolumn{2}{p{1.6cm}<{\centering}|}{ASCAD} & \multicolumn{2}{p{1.6cm}<{\centering}|}{desync50} & \multicolumn{2}{p{1.6cm}<{\centering}|}{desync100} & \multicolumn{2}{p{1.6cm}<{\centering}}{RD}
		\\[2pt]  \cline{2-9}  
		& 0.3 & 0.7 & 0.3 & 0.7 & 0.3 & 0.7 & 0.3 & 0.7 
		\\  \hline
		3000 & 30 & 30 & 195 & 197 & 599 & 303 & 48 & 49 \\  \hline
		50000 & 6 & 7 & 23 & 25 & 44 & 44 & 14 & 16 \\
		\bottomrule[1.5pt]
	\end{tabular}
\end{table}

\end{document}